\documentclass[10pt,aps,prd,floatfix,notitlepage,showpacs,nofootinbib,superscriptaddress,twocolumn,showkeys]{revtex4-1}
\usepackage{amssymb,amsmath}
\usepackage[pagebackref=true,colorlinks,linkcolor=blue,citecolor=magenta]{hyperref}
\usepackage{graphicx}
\begin{document}
\title{Gravitational waves in modified teleparallel theories of gravity}	
	\author{Habib Abedi}
	\email{h.abedi@ut.ac.ir}
	\affiliation{Department of Physics, University of Tehran, North Kargar Ave, Tehran, Iran.}
	\author{Salvatore Capozziello}
	\email{capozziello@na.infn.it}
	\affiliation{Dipartimento di Fisica, Universit\`a di Napoli ''Federico II'', Via Cinthia, I-80126, Napoli, Italy,}
\affiliation{Istituto Nazionale di Fisica Nucleare (INFN), Sez. di Napoli, Via Cinthia, Napoli, Italy,}
\affiliation{Gran Sasso Science Institute, Via F. Crispi 7,  I-67100, L'Aquila, Italy,}
\affiliation{Tomsk State Pedagogical University, ul. Kievskaya, 60, 634061 Tomsk, Russia.}
 
\begin{abstract}
Teleparallel theory of gravity and its modifications have been studied extensively in literature. However, gravitational waves has not been studied enough in the framework of teleparallelism. In the  present study, we discuss gravitational waves in general theories of teleparallel  gravity  containing the  torsion scalar $T$, the boundary term $B$ and a scalar field $\phi$.  The goal is to classify possible new polarizations generalizing results presented in Ref.\cite{Bamba}. We show  that, if  the boundary term is minimally coupled to the torsion scalar and the scalar field, gravitational waves have the same polarization modes  of General Relativity.
	\keywords{Teleparallel gravity; modified gravity; gravitational waves.}
	\pacs{04.50.Kd, 04.30.-w, 98.80.-k}
\end{abstract}
\date{\today}
\maketitle
\section{Introduction}
The observed late expansion of the universe can be described by  either introducing an exotic form of energy (dark energy) or modifying gravity. In this framework, several modifications have been proposed \cite{Nojiri, CapDel, Oiko, Padilla}  and, among them, the possibility to consider teleparallel gravity \cite{RepT}. 
Einstein introduced the idea of teleparallelism soon after General Relativity (GR)~\cite{Einstein}. 
Teleparallel Lagrangian  coincides with Einstein-Hilbert Lagrangian up to a boundary term, \emph{i.e.} $ T=-R+B $, where $T$ is the scalar torsion, $R$ is  the Ricci scalar and $B$ is a boundary term.
Therefore, GR and Teleparallel Equivalent General Relativity (TEGR)  result in the same equations of motion.

However,  difference between them arise in modified Lagrangians, where  scalar fields coupled nonminimally to gravity  or  arbitrary functions of $T$ or $R$ are taken into account
\cite{RepT}.
Such modifications of TEGR  violate the local Lorentz symmetry invariance and  result in six extra degrees of freedom~\cite{Li}.
In a more general case, the Lagrangian can be a function of both $T$ and $R$, \emph{i.e.}  $f(T,R)$~\cite{Myrzakulov,CapoMyr}. This theory can be studied as $f(T,B)$, where $B$ is the boundary term~\cite{Wright2,noetherseb}.

In both GR and TEGR,  gravitational waves (GW) have two independent polarizations, usually denoted as  plus and cross modes.
However,  extra polarizations  appear in  modified theories. 
The perturbation theory in the post-Minkowski limit is a way to study the number of GW polarizations.

The other way is  the Newman-Penrose (NP) formalism~\cite{Eardley1, Eardley2}.
Adopting the  NP formalism in a generic metric theory,  plane GWs have six independent modes of polarization: considering the $z$-direction as the propagation direction of GWs, they are  plus $(+)$, cross $(\times)$, breathing $(b)$, longitudinal $(l)$, vector-$x$ $(x)$ and vector-y $(y)$ modes.
These modes can be described by the  independent NP quantities $ \{ \Psi_2, \Psi_3, \Psi_4, \Phi_{22}\} $, where $\Psi_3$ and $\Psi_4$ are complex and  each one describes two polarization modes.
The extra polarization modes can be used to discriminate among modified theories of gravity beyond GR (see, e.g. Ref.~\cite{ Bogdanos, Calmet}).
 As shown in \cite{Bamba}, GWs  in $f(T)$,  and in its scalar-tensor representation, are equivalent to that in GR and TEGR~\cite{Andrade}.
In $f(R)$ gravity, where the Lagrangian is an arbitrary function of Ricci scalar, three  modes exist ~\cite{Alves1, Capozziello1, Rizwana}. Models $f(R, \Theta)$ and $f(R,\Theta^\phi)$ were also studied in Ref.~\cite{Alves2}, where $\Theta$ and $\Theta^\phi$ are the traces of the energy-momentum tensors of standard matter and of a scalar field, respectively. The Authors studied different form of function $f$, and  have shown that the number of GW-modes depends on the form of it~\cite{Bertolami}.

An important remark is necessary at this point.  Modified theories of gravity  are taken into account  to achieve a comprehensive picture of cosmic dynamics ranging from early inflation, up to large scale structure formation and  current acceleration of the universe \cite{Nojiri, CapDel, Oiko, Padilla}. The approach is aimed to give, in principle, a full geometric description of cosmic history consisting, for example, in extensions of GR, like $f(R)$, or  of TEGR, like $f(T)$. The main task is explaining dynamics by further degrees of freedom of gravitational field (with respect to GR or TEGR) instead of invoking dark components \cite{francaviglia}.  However, to achieve a self-consistent description, further scalar fields could be necessary. For example,  as discussed in \cite{StabileAR}, the flat rotation curve of galaxies is better fitted considering a theory like $f(R,\phi)$,  instead of a pure $f(R)$, because, in such a case, it  is possible to reproduce  the so-called Sanders potential with better precision. In this case, by a conformal transformation, it is  shown that a model like $f(R,\phi)$ is analogue to $f(R,\Box R)$ so that  the scalar field has a straightforward geometric interpretation too. In general, terms like $\Box R$, $\Box^2R$ and so on appear as UV corrections that  have effects also at IR scales (see \cite{CapDel} for a detailed discussion of this topic). In this perspective, further scalar fields, having a geometric  or a matter origin, could be useful to describe coherently cosmic dynamics at any scale. Here, we consider TEGR extensions assuming not only general functions of the torsion scalar $T$, but also boundary terms $B$ and a scalar field $\phi$ that, according to the discussion in \cite{StabileAR}, could be geometrically interpreted. In particular, considering further scalar fields is important for a full classification of GW modes and polarizations.

The present paper is organized as follows.
The field equations of our modified teleparallel theory are derived in Sec.~\ref{Teleparallel}. Sec.~\ref{Nonminimal} is devoted to study GWs in two modifications of teleparallel gravity; first, we study the case of  scalar field  nonminimally coupled  to both the  scalar torsion and the boundary term. Then, we assume a Lagrangian as  a nonlinear function of the scalar torsion and the boundary term. In Sec.~\ref{Kinetic}, we  obtain the number of GW-polarizations   when the scalar field kinetic term is coupled to the scalar torsion. 
We show that,  due to the local Lorentz Invariance  violation, such a  coupling is not viable because of the extra degrees of freedom. In Sec. \ref{Conclusions}, we discuss the results and draw conclusions.
\section{Teleparallel gravity and its extensions}
\label{Teleparallel}
In teleparallel theories, vierbein fields describe gravity.
Considering a set of orthonormal basis in each point of a generic manifold, the metric is given by
\begin{align}
g_{\mu\nu}=\eta_{AB} e^A_{\phantom{A}\mu} e^B_{\phantom{B}\nu},
\end{align}
where $e^A_{\phantom{A}\mu}$ are vierbein fields and $\eta_{AB}$ is the  Minkowski metric.
Then, one can write
$ e^A_{\phantom{A}\mu} e_A^{\phantom{A}\nu}=\delta^\nu_\mu $.
With the rule of absolute transport 
$ \tilde{\nabla}_\mu e_A^{\phantom{A}\nu}=0 $, 
 the Weitzenb\"ock  connection with vanishing Riemann tensor is defined by
\begin{align}
\Gamma^\alpha_{\phantom{\alpha}\nu\mu}:=e_A^{\phantom{A}\alpha}\, \partial_\mu e^A_{\phantom{A}\nu}\,.
\end{align} 
 $ \tilde{\nabla}_\mu$ is the covariant derivative is defined by the Weitzenb\"ck connection.
This connection results in nonvanishing torsion tensor as follows
\begin{align}
T^\alpha_{\phantom{\alpha}\mu\nu}=e_A^{\phantom{A}\alpha}\left(\partial_\mu e^A_{\phantom{A}\nu} - \partial_\nu e^A_{\phantom{A}\mu} \right).
\end{align}
Defining contorsion and superpotential, respectively,
\begin{align}
K^{\mu\nu}&:= -\frac{1}{2} \left(T^{\mu\nu}_{\phantom{\mu\nu}\rho}-T^{\nu\mu}_{\phantom{\nu\mu}\rho}-T_\rho^{\phantom{\rho} \mu \nu}\right),
\\
S_\rho^{\phantom{\rho}\mu\nu}&:=\frac{1}{2} \left(K^{\mu\nu}_{\phantom{\mu\nu}\rho}+\delta^\mu_\rho T^{\alpha \nu}_{\phantom{\alpha \nu}\alpha}-\delta^\nu_\rho T^{\alpha \mu}_{\phantom{\alpha \mu}\alpha}\right),
\end{align}
scalar torsion is
\begin{align} \label{10}
T:=S_\rho^{\phantom{\rho}\mu\nu} T^\rho_{\phantom{\rho}\mu\nu}. 
\end{align}
The scalar torsion~\eqref{10} is related to the Ricci scalar constructed by the Levi-Civita connection as follows
\begin{align}
T=-R+B,
\end{align}
where $B=2\nabla_\mu T_\nu^{\phantom{\nu}\nu\mu}$ is a boundary term in the teleparallel Lagrangian.
If  a scalar field is nonminimally coupled to  the torsion scalar,  the Einstein frame can be recovered by
 considering the boundary term $B$  coupled to the scalar field \cite{Wright1}. Let us now take into account the  following action
\begin{align}
S=\frac{1}{2}\int {\rm d}^4x \, e  \left[  f(T,B, \phi)-  \partial_\mu \phi \, \partial^\mu \phi -2 V(\phi) +2 {\cal L}_{\rm m} \right] , \label{action4}
\end{align}
where $e={\rm det}\left(e^A_{\phantom{A}\nu}\right)=\sqrt{-g}$,  $V(\phi)$ is a generic potential and ${\cal L}_{\rm m}$ is the matter Lagrangian.
The variation of action (\ref{action4}) with respect to the vierbein fields yields the  following field equations
\begin{align} 
&e_A^{\phantom{A} \mu} \square f_B-e_A^{\phantom{A} \nu} \nabla^\mu \nabla_\nu f_B+\frac{1}{2} B f_B e_A^{\phantom{A} \mu} 
+2 \partial_\nu\left(f_B+f_T\right) S_A^{\phantom{A} \nu \mu} \nonumber \\ 
&+2e^{-1} \partial_\nu \left(e S_A^{\phantom{A} \nu \mu}\right) f_T
-2f_T T^\alpha_{\phantom{\alpha} \nu A} S_\alpha^{\phantom{\alpha}\mu \nu}  
\nonumber \\ &-\frac{1}{2}e_A^{\phantom{A} \mu}\left[f- \partial_\alpha \phi \, \partial^\alpha \phi -2 V(\phi)\right]= \Theta_A^{\phantom{A}\mu},  \label{11}
\end{align}
where $\Theta^{\phantom{A}\mu}_A=-\delta{\cal L}_{\rm m} / \delta h^A_{\phantom{A}\mu}$ is the stress-energy tensor of matter.
Eqs.~\eqref{11} in spacetime indices  become
\begin{align}
&-f_T G_{\mu \nu} + \left( g_{\mu \nu} \square - \nabla_\mu \nabla_\nu \right) f_B 
+ \frac{1}{2} ( f_B B+ f_T T -f ) g_{\mu \nu} \nonumber \\ & 	
+2 S^{\phantom{\nu} \alpha}_{\nu \phantom{\alpha} \mu}\, \partial_\alpha (f_T+f_B)
- g_{\mu \nu} \left[ \frac{1}{2}  \partial_\alpha \phi \; \partial^\alpha \phi + V(\phi) \right]
\nonumber \\
&+ \partial_\mu \phi \; \partial_\nu \phi = \Theta_{\mu \nu}, \label{field9}
\end{align}
where we have used 
\begin{equation}
G^\nu_\sigma=-2 \left(e^{-1}\partial_\mu(eS^{\phantom{A}\mu \nu}_A)-T^\rho_{\phantom{\rho}\mu A}S^{\phantom{\rho} \nu \mu}_{\rho} - \frac{1}{4}e^{\phantom{A}\nu}_A T\right) e^A_{\phantom{A} \sigma} .
\end{equation}
The variation of the action~\eqref{action4} with respect to the scalar field results in
\begin{align}
\square  \phi+\frac{1}{2} f^\prime -V^\prime=0 ,
\end{align}
where prime denotes the derivative with respect to the scalar field $\phi$.
In the weak field approximation, the metric can be written as 
\begin{equation}
g_{\mu \nu}=\eta_{\mu\nu}+h_{\mu \nu},
\end{equation}
where $ h_{\mu \nu} $ is small and first order, 
$ {\cal O}\left(h^2\right) \ll 1 $ with respect to the background. Thus, up to first order, one can write
\begin{equation}
e^A_{\phantom{A} \mu}=\delta^A_{\phantom{A}\mu}+h^A_{\phantom{A} \mu}.
\end{equation}
and
\begin{align}
R^{(1)}_{\mu\nu}=&\frac{1}{2}\left(\partial_\rho \partial_\nu h^\rho_{\phantom{\rho}\mu} +\partial^\rho \partial_\mu h_{\nu\rho} - \square h_{\mu \nu} - \partial_\mu \partial_\nu h\right),
\\
R^{(1)}=&\partial_\rho \partial^\mu h^\rho_{\phantom{\rho}\mu} -\square h.
\end{align}
where $h=\eta^{\mu\nu}h_{\mu\nu}$ and $\square=\eta^{\mu\nu} \partial_\mu \partial_\nu$. 
The indices are lowered and raised by the Minkowski background metric $\eta_{\mu\nu}$.
The boundary term $B$ is second order in perturbations; therefore,  up to first order we have
$ R^{(1)}=-T^{(1)} $.
\section{Nonminimal Coupling}
\label{Nonminimal}
\subsection{The role of scalar field}
In order to develop our considerations, we can specify the function in \eqref{action4} as 
\begin{align}
f(T,B,\phi^i)=\left[-1+\xi\, F(\phi)\right]T+ \chi \, E(\phi)\, B \label{5} ,
\end{align}
where $F$ and $E$ are two arbitrary functions of scalar field. For $\xi=0=\chi$ it reduced to TEGR.
Field equations get the following form
\begin{align}
&(-1+\xi F) G_{\mu\nu}+\chi \left(g_{\mu\nu}\square-\nabla_\mu\nabla_\nu \right) E \nonumber \\ &
+2S_{\nu\phantom{\alpha}\mu}^{\phantom{\nu}\alpha} \partial_\alpha \left(\xi F+\chi E\right) -g_{\mu \nu} \left(\frac{1}{2} \partial_\alpha \phi \, \partial^\alpha\phi+V\right) 
\nonumber \\ &+\partial_\mu \phi \, \partial_\nu \phi = \Theta_{\mu \nu}.
\end{align}
At first order we have
\begin{align}
&(-1+\xi F_0) \left(R_{\mu \nu}^{(1)}-\frac{1}{2} \eta_{\mu\nu} R^{(1)}\right) 
+\chi E_0^\prime \left( \eta_{\mu \nu} \partial^2- \partial_\mu \partial_\nu \right) \delta \phi \nonumber \\ &- h_{\mu\nu} V_0
- \eta_{\mu \nu} V^\prime_0 \, \delta \phi = \Theta_{\mu \nu}^{(1)}.\label{9}
\end{align}
Taking the trace of Eq.~\eqref{9}, we get
\begin{align}
-(-1+\xi F_0)R^{(1)}+3\xi E^\prime_0 \square \delta \phi -hV_0 -4V^\prime_0 \, \delta \phi=\Theta^{(1)} .
\end{align}
According to these considerations, we can  define
\begin{align}
\bar{h}_{\mu\nu}=&h_{\mu\nu}-\frac{1}{2} \eta_{\mu \nu}h +\frac{\chi E^\prime_0}{-1+\xi F_0} \eta_{\mu\nu} \, \delta \phi,
\\
\bar{h}=&-h +\frac{4\chi E^\prime_0}{-1+\xi F_0} \delta \phi,
\\
h_{\mu\nu}=&\bar{h}_{\mu\nu}-\frac{1}{2} \eta_{\mu \nu}\bar{h} +\frac{\chi E^\prime_0}{-1+\xi F_0} \eta_{\mu\nu} \, \delta \phi ,
\end{align}
and, in vacuum,  we have
\begin{equation}
\quad\square \bar{h}_{\mu \nu}=0 \quad.
\end{equation}
With the plane wave  ans\"atz, its solution in Fourier space is 
\begin{align}
\bar{h}_{\mu\nu}({\bf k})= A_{\mu \nu}({\bf k}) \, \exp\left(i k^\alpha x_\alpha\right) + {\rm c.c.}  \label{16}
\end{align}
One can assume $\bar{\phi}$ as the minimum of  the potential, \emph{i.e.}
\begin{align}
V\simeq V_0+\frac{1}{2} \gamma \, \left(\delta \phi \right)^2 .
\end{align}
The above scalar field equation, with the choice \eqref{5}, gets the following form
\begin{align}
\square  \phi+\frac{1}{2} \left(\xi T F^\prime+ \chi B E^\prime\right)-V^\prime=0 .
\end{align}
At first order, it becomes
\begin{align}
\square  \delta\phi-\frac{1}{2} \xi F^\prime_0 R^{(1)}-V^{\prime\prime}_0 \delta \phi=0 ,
\end{align}
where we have used $B ={\cal O}(h^2)$ and $T^{(1)}=-R^{(1)}$.
Then, we get
\begin{equation}
\left(\square-m^2\right)\delta \phi=0, \quad
m^2=\frac{2V^{\prime \prime}_0 (-1+\xi F_0)}{2(-1+\xi F_0)-3\xi \chi F^\prime_0 E^\prime_0},,
\end{equation}
where $m^2$ defines an effective mass.
We assumed
$V^\prime_0=0$.
The solution  at first order is then 
\begin{align}
\delta \phi({\bf q})= a({\bf q}) \, \exp\left(iq^\alpha x_\alpha \right) + {\rm c.c.} \label{7}
\end{align}
Let us now consider $z$ as the direction of wave traveling.  Taking $\Omega$ as the angular frequency, we have
\begin{align}
{\bf q}=\left(\Omega,0,0,\sqrt{\Omega^2-m^2}\right),
\end{align}
and the group velocity is
\begin{align}
v_{\rm G} =\frac{\sqrt{\Omega^2-m^2}}{\Omega} .
\end{align}
Assuming the speed $v_G$ constant, we get
\begin{align}
m=\sqrt{(1-v^2_{\rm G})}\Omega.
\end{align}
The effect of gravitational polarization can be studied by the geodesic deviation,
\begin{align}
\ddot{x}_i=-R_{itjt}x_j .
\end{align}
Only the "electric part" of the Riemann tensor, \emph{i.e.} $R_{itjt}$, affects the geodesic deviation. 
In absence of modes that are described by Eq.~\eqref{16}, \emph{i.e.} $\bar{h}_{ij}=0$, we have
\begin{align}
h_{\mu\nu}= \frac{\chi E_0^\prime}{-1+\xi F_0} \eta_{\mu\nu} \, \delta \phi.
\end{align}
Then,  geodesic deviation becomes
\begin{align}
\label{deviation1}
\ddot{x}_i=\frac{\chi E_0^\prime}{2\left(-1+\xi F_0\right)} \left(\eta_{ij} \, \ddot{\delta \phi}+ \left(\delta \phi\right)_{,ij}\right) x_j .
\end{align}
Expressing \eqref{deviation1} in components, one gets 
\begin{align}
\ddot{x}=&-\frac{\chi E_0^\prime \Omega^2}{2\left(-1+\xi F_0\right)} \, \delta \phi \,  x,
\nonumber \\
\ddot{y}=&-\frac{\chi E_0^\prime \Omega^2}{2\left(-1+\xi F_0\right)} \, \delta \phi  \, y, 
\nonumber \\
\ddot{z}=&-\frac{\chi E_0^\prime  m^2}{2\left(-1+\xi F_0\right)} \, \delta \phi \, z .\label{8}
\end{align}
If $\Omega \gg m$, the displacement in longitudinal  direction is smaller than the transverse one,
$ \ddot{z}/z = \left(m/\Omega\right)^2 \, \ddot{x}/x $. 
In very low frequency band, $l$ and $b$ modes can be of the same order.
Considering the weak field limit, we can adopt the  NP formalism. To obtain the independent NP quantities, one can use the solution~\eqref{7}, that is 
\begin{align}
R_{\mu\nu}^{(1)}=\left[\left(-\frac{1}{-1+\xi F_0}+\frac{3}{2} \right) \eta_{\mu\nu} \square + \frac{1}{-1+\xi F_0} \partial_\mu \partial_\nu \right] \chi E^\prime_0 \delta\phi .
\end{align}
Defining a set of tetrads  
$ \left(
{\bf e}_t,{\bf e}_x, {\bf e}_y,{\bf e}_z
\right) $, the null tetrads are
\begin{align}
{\bf k}=& \frac{1}{\sqrt{2}} ({\bf e}_t+{\bf e}_z),&
{\bf l}=& \frac{1}{\sqrt{2}} ({\bf e}_t-{\bf e}_z),
\nonumber\\
{\bf m}=& \frac{1}{\sqrt{2}} ({\bf e}_x+ i{\bf e}_y),&
{\bf \bar{m}}=& \frac{1}{\sqrt{2}} ({\bf e}_x- i{\bf e}_y),
\end{align}
where ${\bf m}$ and ${\bf \bar{m}}$ are complex but  ${\bf l}$ and ${\bf k}$  are real.
The null tetrads satisfy following relations
\begin{align}
-{\bf k}\cdot {\bf l}&= {\bf \bar{m}}\cdot {\bf m}=1,
\nonumber \\
{\bf k} \cdot {\bf l}&= {\bf k}\cdot {\bf \bar{m}}= {\bf l} \cdot {\bf m} = {\bf l} \cdot {\bf \bar{m}} =0.
\end{align}
Then the  non-vanishing NP quantities become
\begin{align}
\Psi_4=&-R_{\bf l\bar{m}l\bar{m}} \sim   \text{$+$ and $\times$ modes}, \\
\Psi_3=&-\frac{1}{2}R_{\bf l \bar{m}} \sim \text{$x$ and $y$ modes}, \\
\Psi_2=& \frac{1}{6} R_{\bf lk}\sim \text{$l$ mode},\\
\Phi_{22}=& -\frac{1}{2} R_{\bf ll} \sim \text{ $b$ mode}.
\end{align}
Then, we have
\begin{align}
\Psi_3=&0, 
\\
\Psi_2=& \frac{\chi E^\prime_0 m^2 a \exp\left(iq_\alpha x^\alpha\right) }{12}\left[\frac{1}{-1+\xi F_0}-\frac{3}{2} \right]
\\
\Phi_{22}=&-\frac{\chi E_0^\prime}{2(-1+\xi F_0)} \exp\left(iq_\alpha x^\alpha\right) \left(q_t-q_z\right)^2
\end{align}
therefore, in general, we have four independent polarizations: $ \times $, $+$, $b$ and $l$ modes. However, the NP formalism can be used for massless waves. Considering $V^{\prime\prime}_0=0$ we have
\begin{align}
\Psi_2=&0=\Psi_3, &
\Psi_4 \neq 0 \neq \Phi_{22}
\end{align}
therefore there exists just three modes: $\times$, $+$ and $b$.
The case in which $\chi=0$ results in $\Phi_{22}=0$, consequently, the two polarization modes of GR remain. Consider that these two polarizations are obtained also in TEGR.  It is worth noticing that the massless scalar field, coupled  with the  boundary term, leads to the breathing mode.
\subsection{The $f(T,B)$ theory}
Let us consider now the  following action
\begin{align}
\label{act}
S=\frac{1}{2} \int {\rm d}^4x \, e \, f(T,B).
\end{align}
The field equations are 
\begin{align}
&-f_T G_{\mu \nu}+\left(g_{\mu \nu } \square - \nabla_\mu \nabla_\nu \right)f_B +\frac{1}{2} g_{\mu \nu} \left(f_B B+f_T T -f\right)
\nonumber \\ & +2  S^{\phantom{\nu}\alpha}_{\nu \phantom{\alpha} \mu} \, \partial_\alpha \left(f_T+f_B\right)=0.
\end{align}
Supposing $f(T,B)$ being an analytic function of $T$ and $B$, one can expand it  as follows
\begin{align}
f(T,B)=&f(T_0,B_0)+f_T(T_0,B_0)\, T+ f_B(T_0,B_0) \, B \nonumber \\ &+ f_{TB}(T_0,B_0)\, TB + \cdots.
\end{align}
Then the  field equations at   first order become
\begin{align} \label{1}
-f_{T_0} G_{\mu\nu}^{(1)}+ f_{T_0 B_0} \left(\eta_{\mu \nu} \square- \partial_\mu \partial_\nu  \right) T^{(1)}=0.
\end{align}
Up to first order we have again $R^{(1)}=-T^{(1)}$. Therefore, we get
\begin{align}
f_{T_0}\left(R^{(1)}_{\mu\nu}-\frac{1}{2}\eta_{\mu \nu} R^{(1)}\right)+f_{T_0 B_0} \left(\eta_{\mu \nu} \square - \partial_\mu \partial_\nu \right)R^{(1)}=0.
\end{align}
Using the  transformation
\begin{align}
h_{\mu \nu}=\bar{h}_{\mu\nu}-\frac{1}{2}\bar{h} \eta_{\mu \nu} -\frac{f_{T_0 B_0}}{f_{T_0}} R^{(1)} \eta_{\mu \nu},
\end{align}
we get
\begin{equation}
\square \bar{h}_{\mu \nu}=0.
\end{equation}
The trace of Eq.\eqref{1} is
\begin{align}  \label{6}
f_{T_0} R^{(1)} -3f_{T_0 B_0} \square R^{(1)}=0.
\end{align}
Then we have
\begin{equation}
\square R^{(1)}+m^2 R^{(1)}=0 ,
\end{equation}
where 
\begin{equation}
 m^2=-\frac{f_{T_0}}{3f_{T_0 B_0}}\,,
 \label{mass}
 \end{equation}
 is the effective mass.
 The solution of this equation is
\begin{align}
R^{(1)}= \hat{R}\left(q^\rho \right) \, \exp\left( i q_\rho x^\rho \right).
\end{align}
One can study different  cases:
\begin{itemize}
	\item
	If	$ f_{T_0 B_0}=0 $ (for example $ F(T)+G(B) $), then, from
	Eq.~\eqref{6}, we get
	\begin{align}
	R^{(1)}=0.
	\end{align}
	\item
	In order to  respect the local Lorentz symmetry invariance, we have to  consider
	$f(T,B)=F(R)$. In this case, the field equations reduce to
	\begin{align}
	F_RG_{\mu \nu}+\left(g_{\mu \nu} \square - \nabla_\mu \nabla_\nu \right) F_R+\frac{1}{2} g_{\mu \nu} \left(F_RR-F\right)=0.
	\end{align}
	By  considering a situation similar to the paper~\cite{2},
	\begin{align}
	F(R)=R+\alpha R^2+ \beta R^3\,,
	\end{align}
	the mass  \eqref{mass} reduces to
	$m^2=-\frac{1}{6\alpha}$ and then results for $F(R)$ gravity can be easily recovered.
\end{itemize} 
Furthermore, the action \eqref{act} can be written as 
\begin{align}
S&
=\frac{1}{2}\int{\rm d}^4x \, e \left[f_{,\phi}T+f_{,\psi}-2U(\phi,\psi)\right] ,
\end{align}
where the new potential is $ 2U(\phi,\psi)=f_{,\phi}+f_{,\psi}\psi -f(\phi,\psi) $.
Varying the action with respect to $\phi$ and $\psi$ by assuming $ f_{,\phi \phi}\neq 0$  and $f_{,\psi \psi} \neq 0$, we get the identifications   $ \phi=T $ and $\psi=B$ that can be used as Lagrange multipliers, that is 
\begin{align}
S&=\frac{1}{2}\int {\rm d}^4x \,e \, \left[f+f_{,\phi}\left(T-\phi\right)+f_{,\psi}\left(B-\psi\right) \right]
\nonumber \\ &=
\frac{1}{2}\int {\rm d}^4x \,e \, \left[f-f_{,\phi} \left({}^{(3)}R+\phi\right)-f_{,\psi}\psi
\right. \nonumber \\ & \left. 
-f_{,\phi} \left(\bar{\Sigma}^{ij} \bar{\Sigma}_{ij} - \bar{\Sigma}^2\right) + (f_{,\phi}+f_{,\psi}){\cal D}_T+ f_{,\psi} {\cal D}_R\right].
\end{align}
Finally, we get
\begin{widetext}
	\begin{align}
	S=  \int  {\rm d}^4x \, N \sqrt{h} \Bigg\{&
	\frac{1}{2} f-\frac{1}{2} f_{,\phi} \left({}^{(3)}R+\phi \right) -\frac{1}{2} f_{,\psi} \psi 
	-\frac{1}{2} f_{,\phi} \left(\bar{\Sigma}^{ij} \bar{\Sigma}_{ij} -\bar{\Sigma}^2\right) +\frac{\bar{\Sigma}}{N} \left(N_j \bar{D}^j f_\psi - f_{,\psi\psi} \dot{\psi}-f_{,\psi \phi}\dot{\phi}\right)
	\nonumber \\ &
	\bar{D}_jf_\psi \, \bar{D}^j \ln N+ h^{ij} T^\alpha_{\phantom{\phi}j \alpha}\, \bar{D}_i(f_{,\phi}+f_{,\psi})- \bar{D}_j (f_{,\phi}+f_{,\psi}) \, \bar{D}^j \ln N+ A^\mu \, \nabla_\mu (f_{,\phi}+f_{,\psi})
	\Bigg\}, \label{2}
	\end{align}
	where
	\begin{align}
	A^\mu=& n^\mu \, \bar{D}_i \omega^i+ \frac{n^\mu}{N} \bar{D}_i\left(N^b B^i_{\phantom{i}b}\right)+ \frac{n^\mu}{2}\left(B^{ij} \, \bar{D}_j \omega_i+ \omega_j \, \bar{D}_i B^{ji}\right)\,.
	\end{align}
\end{widetext}
We have used the integration by parts.
One can simply write the momentum conjugates of degrees of freedom as 
\begin{align}
\pi^\phi=&\frac{\partial S}{\partial \dot{\phi}}=\sqrt{h} \left[-\bar{\Sigma} f_{,\psi \phi} +A^0 N\left(f_{,\phi\phi}+f_{,\psi \phi}\right)\right],
\\
\pi^\psi=&\frac{\partial S}{\partial \dot{\psi}} =\sqrt{h} \left[-\bar{\Sigma} f_{,\psi \psi} +A^0 N \left(f_{,\phi\psi}+f_{,\psi \psi}\right) \right],
\end{align}
\begin{align}
\pi^N=&\frac{\partial S}{\partial \dot{N}}=0,
\quad
\pi^{N^i}=\frac{\partial S}{\partial \dot{N}^i}=0.
\end{align}
The only term that contains time derivative of teleparallel extra degrees of freedom is the second one in the second line of the action; according to our definition of torsion, we have
\begin{widetext}
	\begin{align}
	T^\alpha_{\phantom{\alpha}j \alpha } =T^0_{\phantom{0}j 0}+T^i_{\phantom{i}j i }=&-\frac{1}{N} \partial_j\left(N+N^a \omega_a\right) +\frac{\omega_a}{N} \partial_j \left(N^a+\omega^a N + N^b B^a_{\phantom{a}b} \right) +\frac{1}{N} \partial_0\omega_j -\frac{\omega_a}{N} \partial_0 \left(h^a_{\phantom{a}j}+B^a_{\phantom{a}j}\right) \nonumber \\
	&+ \left(\omega^i+\frac{N^i}{N}\right) \left(\partial_j \omega_i - \partial_i \omega_j\right) + \left( B^i_a+\frac{N^i}{N} \omega_a +h^i_a\right) \left[\partial_i\left(h^a_j+B^a_j\right)-\partial_j\left(h^a_i+B^a_i\right)\right].
	\end{align}
\end{widetext}
Then, using
\begin{align}
\frac{\partial T^\alpha_{\phantom{\alpha}j \alpha}}{\partial \dot{\omega}_k}=&\frac{1}{N} \delta^k_j , \quad
\frac{\partial T^\alpha_{\phantom{\alpha}j \alpha}}{\partial \dot{B}^b_k}= -\frac{\omega_b}{N} \delta^k_j. 
\end{align}
we have
\begin{align}
\pi^{\omega_k}=&\frac{\partial S}{\partial \dot{\omega}_k}= \sqrt{h} h^{ik} \bar{D}_i\left(f_{,\phi}+f_{,\chi}\right) ,
\\
\pi^{B^a_k}=&\frac{\partial S}{\partial \dot{B}^b_k}= -\sqrt{h} h^{ik}\omega_a \bar{D}_i\left(f_{,\phi}+f_{,\chi}\right) .
\end{align}
The momentum conjugate of $h_{ij}$ becomes 
\begin{align}
\pi^{kl}=\frac{\partial S}{\partial \dot{h}_{kl}}=&\frac{\sqrt{h}}{2}\bigg[
-f_{,\phi}\left(\bar{\Sigma}^{kl}-h^{kl} \bar{\Sigma} \right) \nonumber \\ &+\frac{h^{kl}}{N} \left(N_j \bar{D}^j f_{,\psi} - f_{,\psi\psi} \dot{\psi}- f_{,\psi \phi} \dot{\phi}\right) \nonumber \\ &-2 h^{ik} h^{al} \omega_a  \bar{D}_i \left(f_{,\phi}+f_{,\psi}\right)\bigg]\,.
\end{align}
It is worth noticing  that quantities constructed from $h_{ij}$ do not contain any extra degrees of freedom.
Its trace becomes
\begin{align}
\pi=&\frac{\sqrt{h}}{2} \bigg[2f_{,\phi}\bar{\Sigma}+\frac{3}{N} \left(N_j \bar{D}^j f_{,\psi}-f_{,\psi\psi} \dot{\psi} - f_{,\psi\phi} \dot{\phi} \right) \nonumber \\ & - 2 h^{al}\omega_a \bar{D}_i \left(f_{,\phi}+f_{,\psi}\right) \bigg]\,.
\end{align}
In summary, we have classified all possible momenta  related to the degrees of freedom.

\section{Kinetic coupling}
\label{Kinetic}
In action~\eqref{action4} we have considered that gravity couples minimally to kinetic term. In this section,  we study such coupling in view of GW polarizations.
Let us consider the ADM line element,
\begin{align}
{\rm d}s^2=-N^2 \, {\rm d}t^2+ h_{ij}\left( {\rm d}x^i+N^i \, {\rm d}t \right)\, \left( {\rm d}x^j+N^j \, {\rm d}t \right), \label{14}
\end{align}
where $N$, $N^i$‌  and $h_{ij}$‌  are the lapse function,  the shift function  and the metric of three-dimensional space, respectively.
One can write extrinsic curvature as follows
\begin{align}
\bar{\Sigma}_{ij}=\frac{1}{N}\left(\dot{h}_{ij} -\bar{D}_i N_j-\bar{D}_jN_i \right).
\end{align} 
where $ \bar{D}_i $ the 3-Levi-Civita covariant derivative. Then the  Ricci scalar is given by
\begin{align}
R=&{}^{(3)}R+ \bar{\Sigma}^{ij} \bar{\Sigma}_{ij} - \bar{\Sigma}^2 +{\cal D}_R,
\end{align}
where $\bar{\Sigma}=\bar{\Sigma}^{ij} h_{ij}$ is the  trace of the extrinsic curvature and
 \begin{align}
 {\cal D}_R&=\frac{2}{N \sqrt{\gamma} } \partial_t \left(\sqrt{\gamma}\bar{\Sigma} \right) - \frac{2}{N}\bar{D}_i \left(\bar{\Sigma} N^i+\gamma^{ij} \partial_j N \right).
 \end{align}
In GR, $ R $ coupled to $ \partial_\mu \phi \, \partial^\mu \phi $ changes the number of dynamical degrees of freedom (see \cite{schmidt} for a discussion). 
In view of this, let us onsider the action with the following term
\begin{align}
S\supset \int {\rm d}^4x \, \sqrt{-g} R X,
\end{align}
where 
$ X=\frac{1}{2} \partial_\alpha \phi \, \partial^\alpha \phi $ is the  kinetic term. By using the ADM decomposition, we have
\begin{align}
X=-\frac{1}{2N^2} \dot{\phi}^2+\frac{N^i}{N^2} \dot{\phi}\, \partial_i \phi + \frac{1}{2} \left(h^{ij}-\frac{N^i N^j}{N^2}\right) \partial_i \phi \, \partial_j\phi,
\end{align}
and then the action contains the following term
\begin{align}
S \supset& \int {\rm d}^4x \, \sqrt{-g} \left(-\frac{1}{2N^2} \dot{\phi}^2 \right) \, \left[-2\nabla_\mu (\bar{\Sigma}n^\mu)
\right] \nonumber \\
=& \int {\rm d}^4x \,  \frac{\dot{\phi}^2 \dot{\bar{\Sigma}}}{N^2}.
\end{align}
According to this development, the  lapse function is a  dynamical variable.
Therefore,
it is unstable and hence not viable for GWs.
However,  some fine tuned  combination of geometry and scalar field derivatives  exists    which includes $G^{\mu\nu} \partial_\mu \phi \partial_\nu \phi$ where $G^{\mu\nu}$ is the Einstein tensor (see \cite{G1}). 
These 
extra degrees of freedom   cancel out and allow  the models to be stable and avoiding the  Ostrogradskij instability.
In the teleparallel approach, the vierbein fields,  related to the ADM line element~\eqref{14} can be  written as~\cite{TE1}
\begin{align}
e^0_{\phantom{0}\mu}=& (N , {\bf 0}),&
e^a_{\phantom{0}\mu}=& (N^a , h^a_{\phantom{a}i}),
\nonumber \\
e_0^{\phantom{0}\mu}=& (1/N , -N^i/N),&
e_a^{\phantom{a}\mu}=& (0 , h_a^{\phantom{a}i}). \label{15}
\end{align}
The torsion becomes
\begin{align}
T=&-{}^{(3)}R- \bar{\Sigma}^{ij} \bar{\Sigma}_{ij} + \bar{\Sigma}^2 +{\cal D}_T ,
\end{align}
where~\cite{TE1}
\begin{align}
{\cal D}_T&=-\frac{2}{N}\bar{D}_k(N T^{i\phantom{i} k}_{\phantom{i} i}) .
\end{align}
is the boundary terms in  $T$.
Therefore we can split $B$ in a curvature and torsion component, that is
\begin{align}
B={\cal D}_R+{\cal D}_T.
\end{align}
 Clearly ${\cal D}_T$ has no time derivative while  ${\cal D}_R$ contains time derivative of $\bar{\Sigma}$.
This means, in general, that  the boundary term $B$ contains  time derivative.
One can conclude that the coupling of ${\cal D}_R$ or $B$ to the kinetic term will result in instability.

Let us consider now the following action
\begin{align}
S=\int {\rm d}^4x\, e &\left[  R+\frac{1}{2}\partial_\mu \phi \, \partial^\mu \phi -V(\phi)  
\right. \nonumber \\ &\left.+ \frac{1}{2}(\xi T + \chi B) \partial_\mu \phi \, \partial^\mu \phi \right], \label{111}
\end{align}
where $\xi$ and $\chi$ represent  coupling constant to the torsion scalar and the boundary term.  
$\xi+\chi=0$ is the case that  has been  studied in Ref~\cite{G1}, then it was assumed $\chi=\xi=0$. 
However, in action~\eqref{111},  it is enough to consider $\chi=0$, in order to avoid ghost instabilities.
The action we are going to study contains a torsion scalar  nonminimaly coupled to the kinetic term as follows
\begin{align}
S=\int {\rm d}^4x\, e &\left[ -\frac{T}{2}(1+\xi \partial_\mu \phi \, \partial^\mu \phi)+\frac{1}{2}\partial_\mu \phi \, \partial^\mu \phi
\right. \nonumber \\ & 
-V(\phi)+{\cal L}_{\rm m} \bigg]\,. \label{3}
\end{align}
For $\xi=0$, the action~\eqref{3} is equivalent to GR minimally coupled to a scalar field.
Varying with respect to the vierbein fields yields
\begin{align}
&-2\left(1+\xi \partial_\mu \phi \, \partial^\mu \phi \right)\left[e^{-1}\partial_\alpha \big(e S^{\phantom{A} \alpha \nu}_{A} \big)-T^{\rho}_{\phantom{\rho} \beta A} S^{\phantom{\rho} \nu \beta}_\rho \right] 
\nonumber \\ &
+\frac{1}{2} e^{\phantom{A}\nu}_A T- 2 \xi S^{\phantom{A} \alpha \nu}_A \partial_\alpha \big(\partial_\gamma \phi \, \partial^\gamma \phi\big)
\nonumber\\ &
-e^{\phantom{A}\nu}_A \left[\frac{1}{2} \partial_\gamma \phi \, \partial^\gamma \phi - V(\phi)\right] +\partial^\nu \phi \, \partial_A \phi =\Theta^{\phantom{A}\nu}_A . \label{12}
\end{align}
contracting with $e^A_{\phantom{A}\sigma}$, we get
\begin{align}
&G^\nu_\sigma \left(1+\xi \partial_\mu \phi \, \partial^\mu \phi \right) - 2 \xi S^{\phantom{\sigma} \alpha \nu}_\sigma \partial_\alpha \big(\partial_\gamma \phi \, \partial^\gamma \phi\big)
\nonumber\\ 
-&\delta^\nu_{\sigma} \left[\xi T \partial_\gamma \phi \, \partial^\gamma \phi +\frac{1}{2} \partial_\gamma \phi \, \partial^\gamma \phi - V(\phi) \right] 
\nonumber \\ 
+&\partial^\nu \phi \, \partial_\sigma \phi =\Theta^\nu_{\sigma} . \label{4}
\end{align}
The trace of Eq.~\eqref{12} is
\begin{align}
&-2 (1+ \xi \partial_\mu \phi \, \partial^\mu \phi) \left[e^{-1} e^A_{\, \mu} \partial_\alpha (eS^{\, \alpha \nu}_A) +T \right] +2 T 
\nonumber \\ &
- 2 \xi S^{\, \alpha \nu}_\nu \, \partial_\alpha (\partial_\gamma \phi \, \partial^\gamma \phi) -\partial_\gamma\phi \, \partial^\gamma\phi +4 V= \Theta . \label{13}
\end{align}
This modification is not local Lorentz invariant.
Variation of the action with respect to the scalar field also results in
\begin{equation}
\square \phi+V_{,\phi}=\xi \, \partial^\mu\phi \, \partial_\mu T . \label{211}
\end{equation}
Action~\eqref{3} has been  studied in Ref~\cite{Sar1}. In a Friedman-Robertson-Walker  background, we have $T=G_{00}=6H^2$. This implies that the derivative coupling $T \partial_\mu\phi \, \partial^\mu \phi$, on such a  background,   gives the same cosmological evolution as the derivative coupling of the scalar field to the Einstein tensor $G^{\mu \nu} \partial_\nu\phi \, \partial_\mu \phi$. However, beyond background level, they will differ (see also \cite{odigw}).
 Eqs.~\eqref{4} and \eqref{211}, at first order,  results in
\begin{equation}
	\left[G_\sigma^{\nu} -\delta^\nu_\sigma \left(\frac{1}{2} \partial_\gamma\phi \, \partial^\gamma \phi -V(\phi)\right)
	+  \partial^\nu\phi \, \partial_\sigma\phi\right]^{(1)} = \left(\Theta^\nu_\sigma\right)^{(1)},
\end{equation}
and
\begin{equation}
	\left(\square \phi+V_{,\phi}\right)^{(1)}=0.
\end{equation}
These   equations  are exactly the same as equations of motion for a scalar field minimally coupled to the Ricci scalar. Therefore,  the number of GW polarizations are the same as  in the Einstein gravity. 

Under local Lorentz transformation 
$ e^A_{\phantom{A}\mu}=\Lambda^A_{\phantom{A}B}\left(x^\nu \right)\, e^B_{\phantom{B}} $, some quantities of teleparallel gravity are not invariant, \emph{e.g.} torsion tensor becomes
$ T^\alpha_{\phantom{\alpha}\mu\nu}+\Lambda_B^{\phantom{B}A} e_A^{\phantom{A}\alpha} \left(e^C_{\phantom{A}\nu} \partial_\mu- e^C_{\phantom{A}\mu} \partial_\nu\right) \Lambda_C^{\phantom{B}} $.
The infinitesimal local Lorentz transformation is
$ \Lambda^A_{\phantom{A}B}({\bf x})=\left(e^\omega\right)^A_{\phantom{A}B} \simeq \delta^A_{\phantom{A}B}+\omega^A_{\phantom{A}B} $. By breaking this symmetry, six extra degrees of freedom appear~\cite{1}, i.e.
\begin{align}
\omega^0_{\phantom{0}B}=&(0,\omega_B),&
\omega^a_{\phantom{a}B}=&(\omega^a, B^a_{\phantom{a}b}),
\end{align}
where $ B^a_{\phantom{a}b} $ is antisymmetric. Considering these new degrees of freedom, the vierbein fields~\eqref{15}, up to first order, get the following form
\begin{align}
e^0_{\phantom{0}\mu}=& (N+N^a \omega_a , \, \omega_i),\nonumber\\
e^a_{\phantom{0}\mu}=& (N^a+N \omega^a+N^b B^a_{\phantom{a}b} ,\, h^a_{\phantom{a}i}+B^a_{\phantom{a}i}),
\nonumber \\
e_0^{\phantom{0}\mu}=& (1/N , \, -N^i/N-\omega^i),
\nonumber \\ 
e_a^{\phantom{a}\mu}=& (-\omega_a/N ,\, h_a^{\phantom{a}i}+B^i_{\phantom{i}a}+\frac{N^i}{N}\omega_a).
\end{align}
Up to second order in extra degrees of freedom, after some simple calculations, one gets~\cite{1} 
\begin{align}
T=&-{}^{(3)}R+ \bar{\Sigma}^2-\bar{\Sigma}^{ij} \bar{\Sigma}_{ij}+ \frac{2}{N} \bar{D}_i \bar{D}^i N
\nonumber \\ & - \frac{2}{N} \bar{D}_i \left(h^{ij} N T^\alpha_{\phantom{\alpha}j\alpha}\right) 
\nonumber \\
&- 2 \bar{\nabla}_\mu \left[n^\mu \, \bar{D}_i \omega^i + \frac{n^\mu}{N} \bar{D}_i (N^bB^i_{\phantom{i}b})\right] 
\nonumber \\ &-
\bar{\nabla}_\mu \left[n^\mu (B^{ij}\bar{D}_j \omega_i+\omega_j \bar{D}_i B^{ij})\right].
\end{align}
Now, let us consider an action with the following coupling term
\begin{equation}
\int {\rm d}^4x \, e T X.
\end{equation}
The action contains 
\begin{align}
S\supset\int {\rm d}^4x \frac{\dot{\phi}^2}{N^2} \, \partial_0\bigg[&\frac{\bar{D}_i \omega^i}{N}+\frac{\bar{D}(N^b B^i_{\phantom{i}b})}{N^2}+\frac{B^{ij}\bar{D}_j\omega_i}{N}
\nonumber \\ &+\frac{\omega_j \, \bar{D}_iB^{ij}}{N}\bigg].
\end{align}
Therefore, on considering extra degrees of freedom, the torsion scalar coupled into kinetic term results in instability. 
\subsection{$G_{\mu\nu}$ coupled to field derivatives}
Finally, let us consider the action containing the following  term
\begin{align}
G_{\mu\nu} \, \nabla^\mu \phi \, \nabla^\nu \phi\,.
\end{align}
At first order, we get 
\begin{align}
\left(\square -m^2\right) \delta \phi =&0,
\\
R^{(1)}_{\mu\nu}-\frac{1}{2} \eta_{\mu \nu} R^{(1)} =&0.
\end{align}
where the massive term is $m^2=V_{,\phi_0 \phi_0}$.
Therefore, the number of GW polarization is the same as in GR plus a scalar mode related to the presence of the scalar field.
\section{Conclusions}
\label{Conclusions}
The number of GW polarizations depends on the considered   theory of gravity. 
In present work we have studied  GWs in extended   teleparallel gravity  where a boundary term $B$ and a further scalar field $\phi$ are taken into account beside the torsion scalar $T$. The conclusions we reached are the following. There   is no extra polarization in  TEGR  and in  $f(T)$ theory with respect to GR  as already shown in \cite{Bamba}. 
Here we demonstrated that a scalar field, non-minimally coupled to torsion, has only the two polarization of GR plus the scalar mode related to the scalar field itself. However,   new polarizations appear when the scalar field is coupled to the boundary term $B$, beside the standard two modes of  GR.
One can also write the Lagrangian as a function of scalar torsion $T$  and Ricci scalar $R$, however in order to study  GW polarizations,  it is better to decouple the Ricci scalar $R=-T+B$ and then  using  $f(T,B)$.
In $f(T,B)$, extra massless and massive modes arise when the scalar torsion and the boundary term are non-minimally coupled  as in the theory of $f(R)=f(-T+B)$. 
The detection of these extra modes could be a fundamental feature to discriminate between metric and teleparallel approaches (see \cite{RepT} for a discussion). 

 In this perspective,  the GW170817 event \cite{GW_multi} has set important constraints and upper bounds on viable theories of gravity. In fact,  besides the multi-messenger issues, 
the event  provides constraints on the difference between the speed of electromagnetic  and  gravitational waves.  This fact gives a formidable way to fix the mass of  further gravitational  modes which results very light (see \cite{speed1} for details). 
Furthermore   the GW170817 event allows  the investigation of  equivalence principle (through Shapiro delay measurement) and Lorentz invariance. The limits of possible violations of Lorentz invariance  are reduced by the new observations, by up to ten orders of magnitude \cite{speed1}. This fact is extremely relevant to discriminate between metric and teleparallel formulation of gravitational theories. Finally,  GW170817 seems to  exclude some alternatives to GR, including some  scalar-tensor theories like Brans-Dicke gravity, Horava-Lifshitz gravity, and bimetric gravity \cite{lombriser}. 
Considering the present study, the reported data seem in favor of the tensor modes excluding the scalar ones. This means that $f(T)$ gravity, showing the same gravitational modes as GR \cite{Bamba}, should be favored with respect to other teleparallel theories involving further degrees of freedom.
Starting from  these preliminary results, it seems possible a complete classification of modified theories by gravitational waves. However, more events like 
GW170817 are necessary in order to fix precisely gravitational parameters and not giving just upper bounds. In this context, the present study could constitute a sort of paradigm in order to classify gravitational modes and polarizations (see also \cite{Bogdanos,Calmet}). In a forthcoming paper, the comparison with gravitational wave data will be developed in detail.

\begin{acknowledgements}
 SC acknowledges the support of  INFN ({\it iniziative specifiche} TEONGRAV and QGSKY).
 This paper is based upon work from COST action CA15117 (CANTATA), supported by COST (European Cooperation in Science and Technology).
\end{acknowledgements}

\end{document}